\documentstyle[sprocl,epsf]{article}

\begin{document}

\title{Strangeness and Antibaryon Production in Heavy-Ion Collisions}
\author{Fuqiang Wang}
\address{Nuclear Science Division, Lawrence Berkeley National Lab,
Berkeley, CA 94720}

\maketitle

\begin{abstract}
We review the experimental results on strangeness and antibaryon
production in heavy-ion collisions at the AGS and SPS.
We argue that the observed enhancement in kaon production at the AGS
is consistent with hadronic description, while the multistrange baryon 
and antibaryon results at the SPS may need physics beyond hadronic nature.
We call the need for measurements of low energy antilambda-proton 
annihilation cross-section in the modeling of antilambda to
antiproton ratio; large value of this ratio is observed in central 
heavy-ion collisions at the AGS and SPS. 
We speculate the importance of an excitation function measurement of 
the ratio.
\end{abstract}

%
%
\section{Introduction}

Nuclear matter at high energy density has been extensively studied through
high energy heavy-ion collisions.\cite{qm}
The primary goal of these studies is to observe the possible phase 
transition from hadronic matter to quark-gluon plasma (QGP),\cite{qgp}
in which quarks and gluons are deconfined over an extended region.
It is believed that the QGP state existed in the early universe shortly
after the Big Bang, and may still exist in the cores of today's neutron 
stars.\cite{star}

If a QGP is created in a heavy-ion collision, the collision
system will experience the states from deconfined quarks and gluons
to interacting hadrons to, finally, freeze-out  particles
where the measurements are realized. In order to extract
the information about the quark-gluon stage of heavy-ion collisions,
systematic studies of multi-observables at freeze-out as a function
of collision system size and bombarding energy are 
necessary.\cite{signatures}
These observables include strangeness and antibaryon production.

Enhancement in strangeness production in high energy heavy-ion collisions
over elementary nucleon-nucleon (NN) interactions has been proposed as a 
signature of QGP formation.\cite{strangeness,Egg91:QGP}
The idea is that, in the deconfined QGP, strange quark pairs ($s\bar{s}$)
can be copiously produced through gluon-gluon fusion
($gg \rightarrow s\bar{s}$),\cite{Egg91:QGP,Kap86:cs}
while in the hadronic gas, $s\bar{s}$ pairs have to be produced via 
pairs of strange hadrons with higher production thresholds.
Moreover, the time scale of the gluon-gluon fusion process is short, 
on the order of 1--3~fm.\cite{Egg91:QGP}
On the other hand, it is argued that strangeness production rate in a 
chemically equilibrated hadronic gas might be as high as that in a 
QGP.\cite{Kap86:cs,Lee88:prc}
However, the time needed for a hadronic gas system to reach chemical 
equilibrium is significantly longer~\cite{Egg91:QGP,Kap86:cs,Lee88:prc}
than the typical life time of a heavy-ion reaction of the order of 10~fm.

In addition to the enhanced production of strangeness, production of
strange and multistrange antibaryons should be further enhanced in 
QGP~\cite{Lee88:prc,Ko92:prc} under finite baryon density.\footnote{
Experimental measurements indicate that high baryon density is reached
in central heavy-ion collisions at both the AGS and SPS.\cite{baryon}}
The argument can be made in the simple Fermi energy-level picture as the
following. Low energy levels of light quarks ($q$) are already occupied
by the excessive light quarks contributing to the finite baryon density.
When the Fermi energies of light quarks are higher than the bare mass 
of a $s\overline{s}$ pair, $s\overline{s}$ pair production is 
energetically more favorable than that of $q\overline{q}$. 
Hence, the production of nonstrange light antiquarks is suppressed,
resulting in a high $\overline{s}/\overline{q}$ ratio in QGP.

The definition of strangeness enhancement is broad. 
In this article, we use two types of definitions:
(1) enhancement of strangeness production rate and antilambda to 
antiproton ($\overline{\Lambda}/\overline{p}$) ratio in 
central heavy-ion collisions with respect to
peripheral collisions, and/or to isospin weighted NN
interactions at the same energy;
(2) enhancement in the ratio of strange antibaryons, 
$\overline{\Omega}/\overline{\Xi}$ (high strangeness content) with respect
to $\overline{\Xi}/\overline{\Lambda}$ (low strangeness content).

The article is organized in the following way. 
In section~\ref{strangeness}, we discuss strangeness enhancement
at the AGS and SPS. In section~\ref{antibaryon}, we discuss 
strange antibaryon enhancement. 
Since strangeness and strange antibaryons are closely related,
the discussions are necessarily coupled in these two sections. 
In section~\ref{conclusion}, we draw conclusions.

%
%
\section{Strangeness enhancement\label{strangeness}}

At AGS energy, the dominant carriers of strangeness produced in 
heavy-ion collisions are kaons (charged and neutral) and 
hyperons ($\Lambda$ and $\Sigma$'s). 
Charged and neutral kaon yields are approximately equal under charge 
(isospin) symmetry. 
Hyperons are mostly produced through associate production with kaons.
Therefore, charged kaon yields give a fairly good overall scale of
strangeness production at the AGS.

\begin{figure}[hbt]
\centerline{\epsfxsize=2.35in\epsfbox{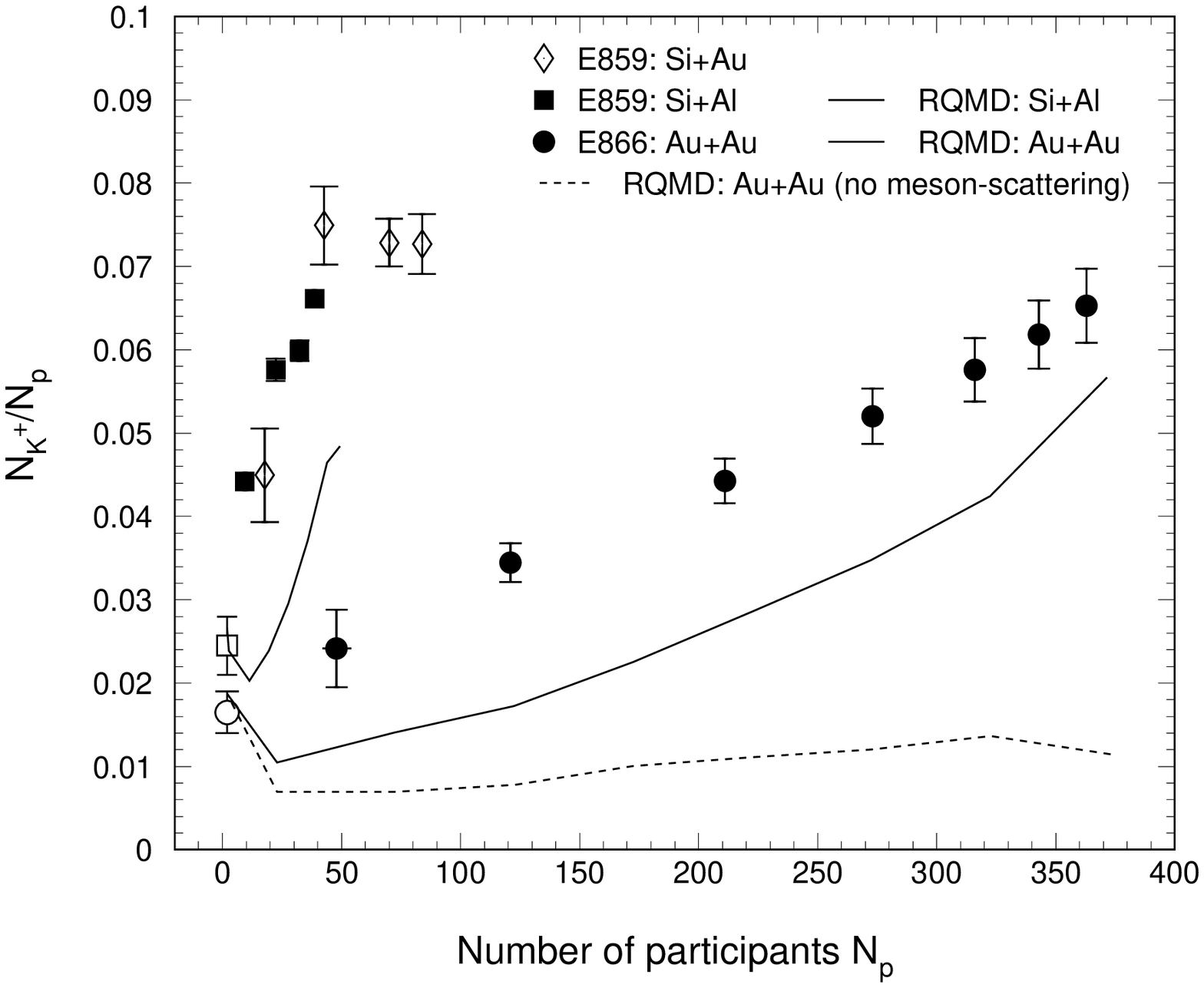}
	    \epsfxsize=2.35in\epsfbox{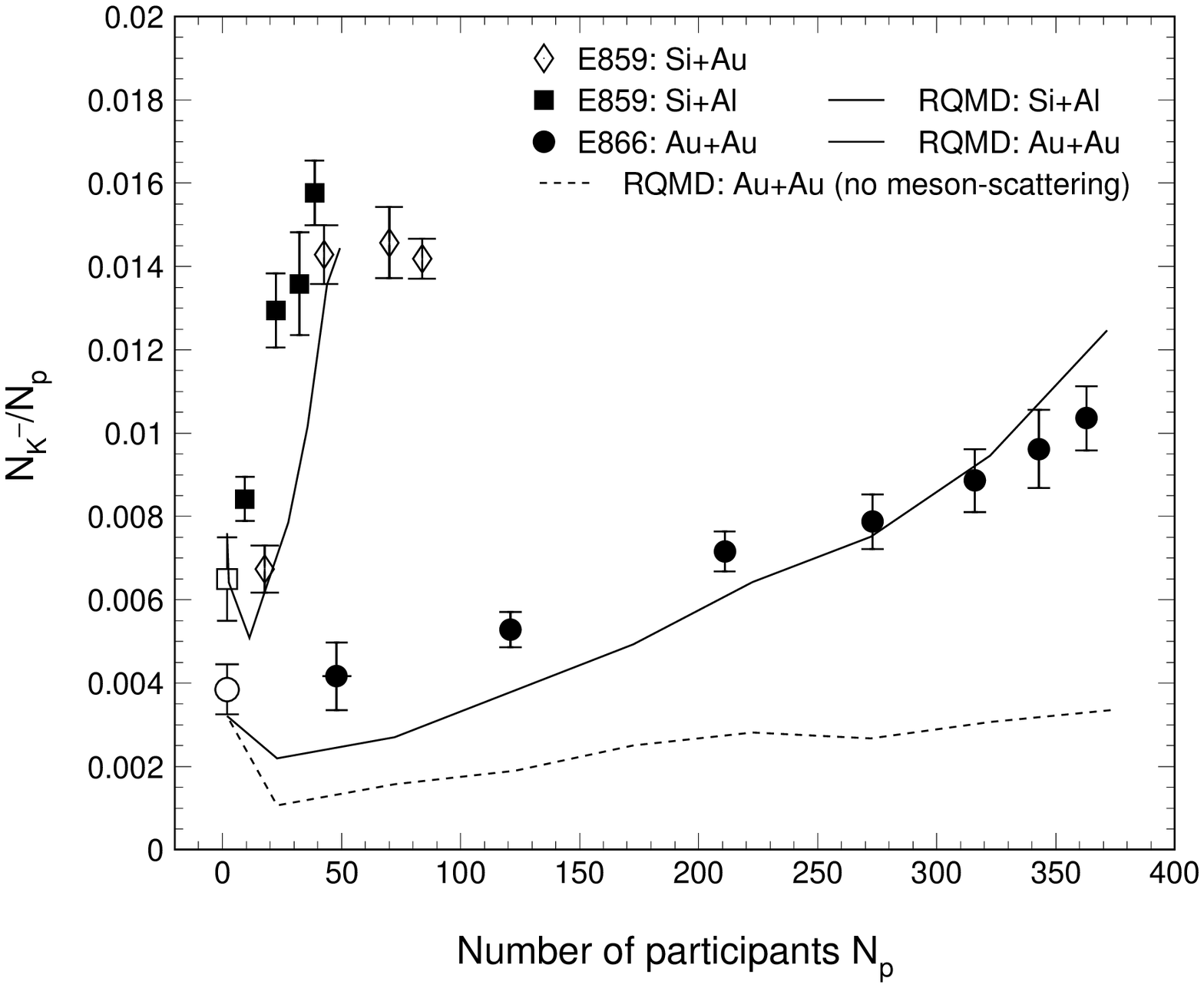}}
\caption{$K^+$ (left panel) and $K^-$ (right panel) yields per
participant as a function of the number of participants ($N_{\rm p}$)
for Si+A ($\sqrt{s}=5.39$~GeV) and Au+Au ($\sqrt{s}=4.74$~GeV) 
collisions at the AGS.$^{11}$
Those for isospin weighted NN interactions at $\sqrt{s}=5.39$~GeV
(open square) and $4.74$~GeV (open dot) are also shown.
{\sc rqmd} results are shown in solid (the default settings) and dashed
curves (meson scattering switched off).
The starting point of each curve is the corresponding {\sc rqmd} result
for isospin weighted NN interaction.
The kaon yields for both data and {\sc rqmd}
include $\phi$ feeddowns, which according to {\sc rqmd}
can be neglected for $K^+$ and increase the $K^-$ yield by 10--15\%
(the default settings).}
\label{fig1}
\end{figure}

Kaon yields are systematically measured by AGS E802/859/866.
Fig.\ref{fig1} shows, in data points, kaon yields per participant
as a function of the number of participants ($N_{\rm p}$) in Si+A and 
Au+Au collisions.\cite{Ahl99:kaons}
$N_{\rm p}$ is a good indication of collision centrality under
the participant-spectator picture of high energy heavy-ion collisions.
The data show that kaon production rate steadily increases with collision
centrality in Si+Al and Au+Au collisions.
For comparison, kaon yields per participant for isospin weighted
NN interactions at the corresponding energies are shown in
Fig.\ref{fig1} as the open circles and squares.
The enhancement factor -- the ratio of kaon yields per participant
in heavy-ion collisions over the same energy NN interactions -- 
is higher in Au+Au central collisions than Si+A for both 
$K^+$ and $K^-$.\cite{Ahl99:kaons}

The difference between Si+A and Au+Au at the same number of participants
(say $N_{\rm p}\sim 50$) is qualitatively consistent with the different
collision geometries under a binary collision model for kaon production, 
however, cannot be quantitatively explained by Glauber-type calculations
of the average number of binary NN collisions.\cite{Ahl99:kaons}

From a classical Glauber-type picture, in which each nucleon collides
multiple times with others in a heavy-ion collision, 
with each collision having certain probability producing kaons,
one would expect higher kaon production rate in heavy-ion collisions 
than NN.
To check this expectation, Fig.\ref{fig1} shows, 
in the dashed curves, results of {\sc rqmd} calculations for Au+Au 
collisions with meson-baryon and meson-meson interactions switched off.
One observes slight increase in kaon production rate from peripheral to
central collisions.
However, in contrast to the expectation, the calculated kaon production
rate in heavy-ion collisions is lower than the corresponding NN value. 
This is probably due to the net effect of the following:
(1) not all the initial nucleons collide at least once at full energy;
(2) subsequent NN collisions have less than the full energy;
(3) AGS energy is near kaon production threshold where kaon production rate
changes rapidly with energy in elementary p+p interactions.\cite{Ros75:pp}

By including secondary meson-baryon and meson-meson interactions 
(the default settings), {\sc rqmd} is able to describe the qualitative
features seen in the data. The results of the default calculations
are shown in Fig.\ref{fig1} as solid curves.
Within the {\sc rqmd} model, both baryon-baryon and meson-induced
interactions increase kaon production rate. 
However, as seen from comparison between the dashed and 
solid curve (for Au+Au) in Fig.\ref{fig1}, 
the meson-induced interactions dominate for kaon production,
especially in central collisions.\cite{Sor91:meson}
Since {\sc rqmd} is a hadronic model,\footnote{The non-hadronic part
of {\sc rqmd}, string formation and fragmentation, does not contribute
significantly at AGS energy.}
and it describes the general features of the data, it is fair to conclude
that the observed kaon enhancement is consistent with hadronic
physics. However, we note that subtle difference exists
between the data and the default {\sc rqmd} results.

As seen from Fig.\ref{fig1}, 
the centrality dependences of $K^+$ and $K^-$ yields are
quite similar. The similarity is surprising because $K^+$ and $K^-$
are thought to be produced through different mechanisms: $K^-$ through
pair production, and $K^+$ through both pair and associate production.

Kaon enhancement in heavy-ion collisions over NN
is also observed at the SPS.\cite{sps_kaons}
Recent preliminary results from NA49~\cite{sikler} show a similar 
centrality dependence of kaon yields as observed at the AGS.

\begin{figure}[hbt]
\centerline{\epsfxsize=2.8in\epsfbox{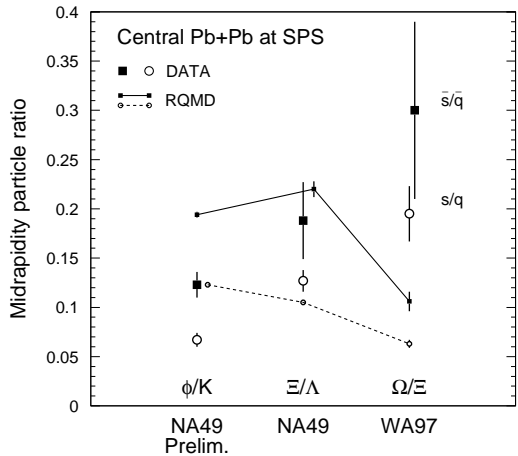}}
\caption{Midrapidity particle ratios 
($\phi/K^+, \Xi^-/\Lambda, \Omega^-/\Xi^-$ in open curcles) 
and antiparticle ratios ($\phi/K^-, \overline{\Xi^-}/\overline{\Lambda}, 
\overline{\Omega^-}/\overline{\Xi^-}$ in filled squares) measured by 
NA49 (5\% centrality)$^{18,19}$ and WA97 (30\% centrality)$^{17}$
in Pb+Pb collisions at the SPS. The {\sc rqmd} results ($b<4$~fm)
are shown in small points connected by dashed (particle ratios) 
and solid line (antiparticle ratios).}
\label{fig2}
\end{figure}

Data on production of the more exotic multistrange baryons and 
antibaryons are also available at the SPS. 
WA97 results~\cite{And98:wa97_plb,Kra98:wa97_qm97} show that the 
$\Lambda, \Xi$ and $\Omega$ midrapidity yields in the 30\% most central
Pb+Pb collisions are larger than those expected from p+Pb by a 
wounded-nucleon scaling. The effect is the largest for $\Omega$, 
following the trend $\Omega>\Xi>\Lambda$. 
More interestingly, the $\Omega/\Xi$ ratio is 
larger than $\Xi/\Lambda$.\cite{Kra98:wa97_qm97}
NA49 reported results on midrapidity $\Xi$ production in the 5\% most
central Pb+Pb collisions,\cite{App98:na49_xi} consistent with the 
WA97 results taken into account the different centralities.
Fig.\ref{fig2} shows the $\Xi/\Lambda$ and $\phi/K$ ratios 
from NA49 (5\% centrality)~\cite{App98:na49_xi,na49_phi}, and the
$\Omega/\Xi$ ratio from WA97 (30\% centrality)~\cite{Kra98:wa97_qm97}
in Pb+Pb collisions.
In the simple quark counting model, the particle ratios
($\phi/K^+, \Xi^-/\Lambda, \Omega^-/\Xi^-$) present the ratio of $s/q$,
and the antiparticle ratios ($\phi/K^-, \overline{\Xi^-}/\overline{\Lambda},
\overline{\Omega^-}/\overline{\Xi^-}$) present the ratio of 
$\overline{s}/\overline{q}$. In Fig.\ref{fig2}, one observes
a hierarchy in the ratios: $\Omega/\Xi>\Xi/\Lambda>\phi/K$, and that
the antiparticle ratios are systematically larger than the particle ratios,
indicating non-vanishing baryon density at midrapidity in Pb+Pb collisions.

{\sc rqmd} predictions for central Pb+Pb collisions 
(impact parameter $b<4$~fm) are shown in Fig.\ref{fig2} 
as the small points connected by the lines. 
{\sc rqmd} agrees with the $\Xi/\Lambda$ data, 
but overpredicts $\phi/K$ by a factor of 1.5--2, while 
underpredicts $\Omega/\Xi$ by a factor of 2.
The discrepency in $\Omega/\Xi$ between the data and {\sc rqmd} may due to
an underestimate of the string fragmentation probability into $\Omega$
in $e^+e^-$, used in {\sc rqmd}.\cite{sorge}
An even larger discrepency is seen between data and
{\sc urqmd} model prediction.\cite{bass}
Presumably, the better agreement between data and {\sc rqmd} than
{\sc urqmd} is due to the novel ``rope'' mechanism,\cite{rqmd_rope}
implemented in {\sc rqmd} but not in {\sc urqmd}. However,
the ``rope'' mechanism is clearly beyond the scope of hadronic physics.

\section{Antibaryon enhancement\label{antibaryon}}

As discussed in the introduction, enhancement in strange antibaryon 
production is a possible signature for QGP formation.
Preliminary results from AGS E859 indicate a large
$\overline{\Lambda}/\overline{p}$ ratio in Si+Au central collisions
near midrapidity.\cite{yuedong}
Direct measurement of $\overline{\Lambda}$ production in Au+Au collisions
at the AGS was not available until recently.\cite{e917}
However, the $\overline{\Lambda}/\overline{p}$ ratio can be inferred
from $\overline{p}$ measurements in the midrapidity region by 
E864~\cite{Arm97:prl,Arm97:prc} and E878~\cite{e878} at $p_T\approx 0$ 
with different acceptances for $\overline{p}$ from weak decays (
$\overline{\Lambda}+\overline{\Sigma^0}$ and $\overline{\Sigma^+}$),
provided that the experimental acceptances and systematics
are well understood. The obtained
$(\overline{\Lambda}+\overline{\Sigma^0}+1.1\overline{\Sigma^+})/\overline{p}$
ratio increases strongely with 
collision centrality, with a most probable value of 3.5 in the 10\% 
most central collisions.\cite{Arm97:prc}

\begin{figure}[hbt]
\centerline{\epsfxsize=2.8in\epsfbox{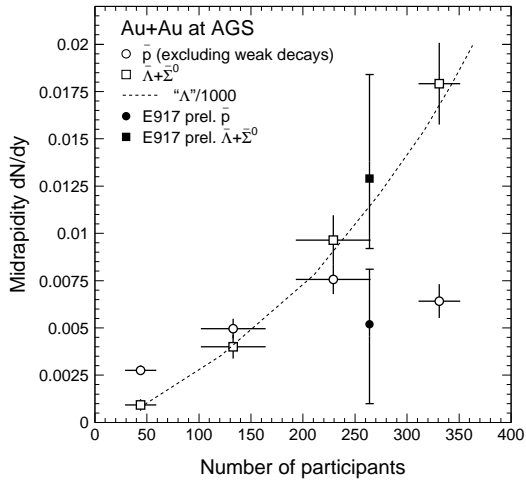}}
\caption{Midrapidity yields of $\overline{\Lambda}+\overline{\Sigma^0}$
and $\overline{p}$ (excluding weak decay products) as a function of 
collision centrality in Au+Au collisions at the AGS, deduced from
midrapidity $\overline{p}$ measurements by E866$^{28}$ and the most probable
$(\overline{\Lambda}+\overline{\Sigma^0}+1.1\overline{\Sigma^+})/\overline{p}$
ratio by E864$^{26}$, assuming 
$\overline{\Sigma^0}=\overline{\Sigma^+}=\frac{1}{3}\overline{\Lambda}$
under isospin consideration.
``$\Lambda$''=$K^+-K^-$ by the constrain of overall strangeness neutrality
and the assumption of $K^+-K^-=K^0-\overline{K^0}$ and $\Lambda=\Sigma$.}
\label{fig3}
\end{figure}

Assuming $\overline{\Sigma}$ yield of each sign is 1/3 of 
$\overline{\Lambda}$ yield under isospin consideration,\footnote{
The assumption does not account for the mass difference between
$\overline{\Lambda}$ and $\overline{\Sigma}$, which may yield a 
deviation from 1/3 in a thermal model argument. 
However, the assumption of 1/3 is fairly consistent 
with {\sc {\sc rqmd}} and {\sc venus} model calculations.}
the $\overline{\Lambda}/\overline{p}$ ratio can be deduced.
Further using the E866 measurements of $\overline{p}$'s which include 
60\% weak decay $\overline{p}$'s,\cite{Ahl98:prl} the yields of 
$\overline{p}$ and $\overline{\Lambda}$ can be obtained, the results
of which are shown in Fig.\ref{fig3} against collision centrality.
The recent direct measurements of $\overline{p}$ and $\overline{\Lambda}$
yields by E917~\cite{e917} are plotted as filled points.
They agree with the obtained systematics, although suffering from large 
statistical errors.

As seen from Fig.\ref{fig3}, $\overline{p}$
yield increases from peripheral to medium central collisions, 
and saturates (or even slightly decreases) in central collisions.
This is consistent with the picture that more $\overline{p}$'s are
absorbed in the nuclear medium in central collisions with higher 
baryon density and/or larger collision zone. 
However, the continuous increase of the $\overline{\Lambda}$ yield
is surprising. In the additive quark model, $\overline{\Lambda}N$
annihilation cross section is 2/3 of $\overline{p}N$'s,
therefore, a strong absorption of $\overline{\Lambda}$
should also occur. 

Because of the lack of $\overline{\Lambda}N$ annihilation cross-section
data at the relavent low energy ($\sim 1$~GeV), hadronic models have to
make assumptions for the cross-seciton to calculate 
$\overline{\Lambda}/\overline{p}$.
Cascade calculations by Wang {\em et al.},\cite{Wan98:ratios}
using various parameterizations for the $\overline{\Lambda}N$ 
annihilation cross-section,
fail to describe the observed $\overline{\Lambda}/\overline{p}$
in central collisions, while giving reasonable agreement to the 
peripheral collision data.

It would be interesting to compare the centrality dependences of 
$\overline{\Lambda}$ and $\Lambda$ yields.
Since the $\Lambda$ yield centrality dependence at the AGS
is presently not available, we use the difference in $K^+$ and $K^-$ 
yields to reflect the $\Lambda$ yield, exploiting the constrain of
overall strangeness neutrality, and further assuming 
$K^+-K^-=K^0-\overline{K^0}$ and $\Lambda=\Sigma$ 
under isospin consideration.
The results are shown in Fig.\ref{fig3} as the dashed curve.
It is interesting to note that the centrality depedences of the
$\Lambda$ and $\overline{\Lambda}$ yields are very similiar,
a feature also observed for $K^+$ and $K^-$ yields.

\begin{figure}[hbt]
\centerline{\epsfxsize=2.6in\epsfbox{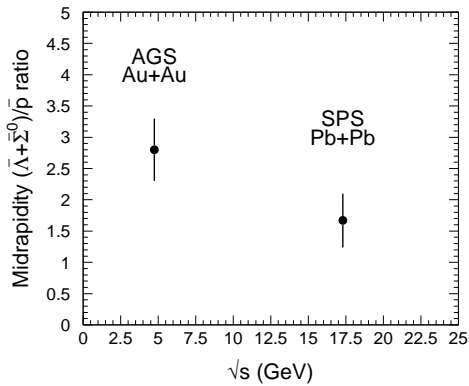}}
\caption{Midrapidity ($\overline{\Lambda}+\overline{\Sigma^0})/\overline{p}$
ratios in central collisions of Au+Au at the AGS and Pb+Pb at the SPS.}
\label{fig4}
\end{figure}

From the centrality dependence of $\overline{\Lambda}/\overline{p}$,
it is hard to disentangle the two different effects, QGP formation and
strong medium absorption, both giving an increasing
$\overline{\Lambda}/\overline{p}$ with centrality.
It is therefore attempting to examine the ratio as a function of
collision energy, which is shown in Fig.\ref{fig4}.\footnote{
The ratio for Pb+Pb at the SPS is derived from several 
measurements.\cite{Kra98:wa97_qm97,sps_ratio}}
The two effects would give different dependences: at high energy,
QGP formation would result in a high ratio, while strong absorption
would result in a low ratio because the baryon
density is lower at high energy.\cite{baryon,wang_delphi}
Therefore, an increase of $\overline{\Lambda}/\overline{p}$
with collision energy would indicate QGP formation or other new physics.
From the current data shown in Fig.\ref{fig4}, we cannot 
draw conclusions. Data at other energies are therefore highly desirable.

The $\overline{\Omega}/\overline{\Xi}$ and 
$\overline{\Xi}/\overline{\Lambda}$ ratios discussed in 
section~\ref{strangeness} are significantly lower than
$\overline{\Lambda}/\overline{p}$ shown in Fig.\ref{fig4} for
Pb+Pb central collisions.
This points to the speculation that medium absorption of strange 
antibaryons may be very different from that of antiprotons.
Again, measurements of low energy $\overline{\Lambda}N$ 
annihilation cross-sections are indispensable.

\section{Conclusions\label{conclusion}}

We reviewed the AGS and SPS data on strangeness and antibaryon production.
We draw the following conclusions:
\begin{itemize}
\item[--] Enhancement in charged kaon production is observed in central 
	heavy-ion collisions at the AGS with respect to peripheral 
	collisions and isospin weighted nucleon-nucleon interactions. 
	The enhancement can be qualitatively explained by secondary 
	particle scattering within the {\sc rqmd} model.
	However, subtle difference exists between the data and {\sc rqmd}.
\item[--] A hierarchy of particle ratios, $\Omega/\Xi>\Xi/\Lambda>\phi/K$,
	is observed at midrapidity in Pb+Pb central collisions at the SPS. 
	The antiparticle ratios are systematically larger than 
	the particle ratios.
	{\sc rqmd} with string and rope formation and fragmentation,
	which is clearly beyond hadronic physics, may barely describe
	the data.
\item[--] Large $\overline{\Lambda}/\overline{p}$ ratios are observed in
	central heavy-ion collisions at the AGS and SPS.
	Strong increase of the ratio from peripheral to central 
	collisions is indirectly observed at the AGS.
	Hadronic cascade calculations, with certain assumptions for
	the $\overline{\Lambda}N$ annihilation cross-section, 
	fail to describe the ratio in central collisions.
\item[--] To fully understand the $\overline{\Lambda}/\overline{p}$ 
	ratio in heavy-ion collisions, measurements of low energy 
	$\overline{\Lambda}N$ annihilation cross-sections are needed.
	On the other hand, an excitation function measurement of
	$\overline{\Lambda}/\overline{p}$ may provide a new avenue
	in the search for QGP.
\end{itemize}

%
\section*{Acknowledgments}

I am especially grateful to Professor R.K.~Seto for 
inviting me to review the experimental results.
I would like to thank Drs. A.M.~Poskanzer, H.G.~Ritter, N.~Xu,
and other members of the LBNL/RNC group for fruitful discussions.
This work was supported by the U.S. Department of Energy
under contract DE-AC03-76SF00098.

%
%
\section*{References}

\end{document}